\documentstyle[amssymb,aps]{revtex}

\begin{document}
\title{Method of convex rigid frames and applications in studies of multipartite
quNit pure-states }
\author{Zai-Zhe Zhong}
\address{Department of Physics, Liaoning Normal University, Dalian 116029, Liaoning,\\
China. Email: zhongzaizheh@hotmail.com}
\maketitle

\begin{abstract}
In this Letter we suggest a method of convex rigid frames in the studies of
the multipartite quNit pure-states. We illustrate what are the convex rigid
frames and what is the method of convex rigid frames. As the applications we
use this method to solve some basic problems and give some new results
(three theorems): The problem of the partial separability of the
multipartite quNit pure-states and its geometric explanation; The problem of
the classification of the multipartite quNit pure-states, and give a perfect
explanation of the local unitary transformations; Thirdly, we discuss the
invariants of classes and give a possible physical explanation.

PACC numbers: 03.67.Mn, 03.65.Ud, 03.67.Hk
\end{abstract}

It is known that in quantum mechanics and quantum information, contrast the
case of the bipartite quantum systems with the case of studies of the
multipartite quantum systems, the latter is even more difficult. For
instance, for the general multipartite quantum systems the problems of the
criteria of various separability, of the entanglement measures, of the
classification and invariants, etc., all are not solved better as yet. In
the studies of the multipartite quantum pure-states, generally we always use
the traditional way, i.e. we discuss the state vectors or the density
matrixes in the Hilbert space, etc.. However, sometimes this way is not
quite effective, especially for some problems the results always are short
of an explicit or geometric explanation. This urges us to find some
non-traditional ways in quantum mechanics and quantum information. The
purpose in this Letter is just to discuss some problems in this respect.

In this Letter, first we illustrate what is a convex rigid frame, and we
suggest a new way, we call it the `{\bf method of convex rigid frames}' (see
below), which associates a multipartite quNit pure-state to a convex
polyhedron and its a point in the Hilbert-Schmidt (H-S) space (On the real
number field all Hermitian operators acting upon a Hilbert space form a
linear space, it is called the Hilbert-Schmidt space). Sometimes, this
method is more effective. As some applications, in this Letter we use this
method to study three basic problems and give some new results (three
theorems): The first is the problem of the so-called partial separability of
the multipartite quNit pure-states and its a perfect geometric explanation;
Secondly we discuss the problem of the classification of the multipartite
quNit pure-states, and give a perfect geometric explanation of the local
unitary transformations; Thirdly, we discuss the invariants of classes and
give a possible physical explanation{\em .}

Sometimes, we call a vector (operator) in the H-S space a `point'. In this
Letter, the operators (vectors, points) considered by us all are the density
matrixes. In the H-S space$,$ the interior product between two vectors $A$
and $B$ is defined as[1] $<A,B>$=$tr\left( A^{\dagger }B\right) ,$ the
modulus of a vector $A$ is defined by $\left\| A\right\| =\sqrt{<A,A>}=\sqrt{%
tr\left( A^{\dagger }A\right) }.$ The distance $d\left( A,B\right) $ between
two points $A$ and $B$ is defined by $d\left( A,B\right) =\left\|
A-B\right\| $ . In the H-S space$,$ if a $n$-convex polyhedron {\bf C}$_n$
has $n$ vertexes $\sigma _i\left( i=1,\cdots ,n\right) $, then the convex
sum $\sigma =$ $\sum\limits_{i=1}^n\lambda _i\sigma _i$ $\left( 0\leqslant
\lambda _i\leqslant 1,\sum\limits_{i=1}^n\lambda _i=1\right) $ denotes a
point in {\bf C}$_n$, we label this point $\sigma $ by $\left( \lambda
_i\right) \equiv \left( \lambda _1,\cdots ,\lambda _n\right) $. We denote
the set of above $n$ vertexes $\left\{ \sigma _i\right\} $and the fixed
point $\sigma $ together a symbol $\left\{ \left( \sigma _i\right) ,\left(
\lambda _i\right) \right\} .$ In this Letter, for the study of the $M$%
-partite quNit pure-states, every related convex polyhedron {\bf C}$_n$ and
the corresponding point $\sigma $ only can moved as a rigid body as in the
classical mechanics, so we call it a `$n$-convex rigid frame ($n$-CRF)',
simply read it the symbol $CRF=\left\{ \left( \sigma _i\right) ,\left(
\lambda _i\right) \right\} $.

{\it Definition 1}{\bf . }Two n-convex rigid frames $CRF$=$\left\{ \left(
\sigma _i\right) ,\left( \lambda _i\right) \right\} $\ and $CRF^{\prime
}=\left\{ \left( \sigma _i^{\prime }\right) ,\left( \lambda _i^{\prime
}\right) \right\} $\ are called to be identical, if $d\left( \sigma
_i,\sigma _j\right) =d\left( \sigma _i^{\prime },\sigma _j^{\prime }\right) $%
\ and $\lambda _i=\lambda _i^{\prime }$ for any $i,j=1,\cdots ,n$.\ In this
case we call the process $CRF\longrightarrow CRF^{\prime }$ a `motion from $%
CRF$\ to $CRF^{\prime }$'{\it .}

Obviously, this{\it \ }identical relation is an equivalence relation,
therefore all $n$-CRFs can be classified by this identical relation.

Now, we consider a multipartite quantum system $H=\otimes _{i=1}^MH_i$\ with 
$M$ parties, all local Hilbert spaces $H_i$ have the same dimension $N,$ then%
$\ $the total dimensionality of $H$ is $N^M$. Under the standard natural
basis $\left\{ \mid i_1\cdots i_M>\right\} \left( i_k=0,1,\cdots ,N-1\text{
and }k=1,\cdots ,M\right) ,$ a normalized $M$-partite quNit state vector $%
\mid \Psi >$ $\in H$ is in form as 
\begin{equation}
\mid \Psi >=\sum_{i_1,\cdots ,i_M=0}^{N-1}c_{i_1i_2\cdots i_M}\mid i_1\cdots
i_M>,\;\;c_{i_1i_2\cdots i_M}\in {\Bbb C}^1,\;\sum_{i_1,\cdots
,i_M=0}^{N-1}\left| c_{i_1i_2\cdots i_M}\right| ^2=1
\end{equation}
In the following, we denote the set of all $M${\bf -}partite quNit
pure-state density matrixes $\rho =\mid \Psi ><\Psi \mid $ by the symbol $%
{\Bbb P}_{M\times N},$ then ${\Bbb P}_{M\times N}$ is a set of points in the 
$N^{2M}$-dimensional H-S space. For a given $\rho =\mid \Psi ><\Psi \mid $,
by the following way we at once can obtain a set of CRFs. In the following, $%
{\Bbb Z}_M$ denotes the integer set ${\Bbb Z}_M=\left\{ 1,\cdots ,M\right\}
, $ and $\left( r\right) _P$ denotes a non-null, proper and naturally
ordered subset in ${\Bbb Z}_M,$ $\left( r\right) _P\subset {\Bbb Z}_M,\left(
r\right) _P\equiv \left\{ r_1,\cdots ,r_P\right\} ,$ where $1\leqslant
P\leqslant M-1$, $\;r_1<\cdots <r_P$, and we denote the set $\left[
i_{\left( r\right) _P}\right] \equiv \left\{ i_{r_1},\cdots ,i_{r_P}\right\}
\left( i_{r_1},\cdots ,i_{r_P}=0,\cdots ,N-1\right) .$ Now, for a $\mid \Psi
>$ as in Eq.($1$) and any fixed set $\left[ i_{\left( r\right) _P}\right] ,$
we define a $\left( M-P\right) $-partite quNit pure-state $\mid \Psi \left[
i_{\left( r\right) _P}\right] >$ by 
\begin{equation}
\mid \Psi \left[ i_{\left( r\right) _P}\right] >=\sum_{\;i_{s_1},\cdots
,i_{s_{M-P}}=0,\cdots \text{ ,N-1 for all }s_1,\cdots ,s_{M-P}\notin \left(
r\right) _P}c_{i_1\cdots i_M}\mid i_1\cdots i_M>
\end{equation}
i.e. for $\mid \Psi \left[ i_{\left( r\right) _P}\right] >,$ the indexes $%
i_{r_1},\cdots ,i_{r_P}$ are fixed, sum up only for the othrs, $%
i_{s_1},\cdots ,i_{s_{M-P}}$. Notice that $\mid \Psi \left[ i_{\left(
r\right) _P}\right] >,$ generally, is not normalized, we make the
normalization 
\begin{equation}
\mid \varphi \left[ i_{\left( r\right) _P}\right] >=\left( \eta _{\left[
i_{\left( r\right) _P}\right] }\left( \rho \right) \right) ^{-1}\mid \Psi
\left[ i_{\left( r\right) _P}\right] >,\;\eta _{\left[ i_{\left( r\right)
_P}\right] }\left( \rho \right) =\sqrt{\sum_{\;i_{s_1},\cdots
,i_{s_{M-P}}=0,\cdots \text{ ,N-1 for all }s_1,\cdots ,s_{M-P}\notin \left(
r\right) _P}\left| c_{i_1\cdots i_N}\right| ^2}
\end{equation}
where $\eta _{\left[ i_{\left( r\right) _P}\right] }\left( \rho \right) $ is
the normalization factor. We write the pure-state density matrix $\sigma
_{\left[ i_{\left( r\right) _P}\right] }\left( \rho \right) \equiv \mid
\varphi \left[ i_{\left( r\right) _P}\right] ><\varphi \left[ i_{\left(
r\right) _P}\right] \mid ,$ of them (for all possible $\left[ i_{\left(
r\right) _P}\right] $) the total is $N^P.$

Now for every pure-state density matrix $\rho =\mid \Psi ><\Psi \mid $, from
the normalization condition of $\mid \Psi >$ we have 
\begin{equation}
\sum\limits_{\text{for all possible }\left( r\right) _P}\eta _{\left[
i_{\left( r\right) _P}\right] }^2\left( \rho \right) =1
\end{equation}
then $\sigma _{_{\left( r\right) _P}}\left( \rho \right) =\sum\limits_{\text{%
for all possible }\left( r\right) _P}\lambda _{\left[ i_{\left( r\right)
_P}\right] }\left( \rho \right) \sigma _{\left[ i_{\left( r\right)
_P}\right] }\left( \rho \right) $ is a point in the $N^P$-convex polyhedron
with vertexes $\left\{ \sigma \left[ i_{\left( r\right) _P}\right] \right\}
, $where $\lambda _{\left[ i_{\left( r\right) _P}\right] }=\eta _{\left[
i_{\left( r\right) _P}\right] }^2\left( \rho \right) .$ Thus, for every
pur-state density matrix $\rho $ we always give a corresponding $N^P$-CRF as 
\begin{equation}
CRF_{_{\left( r\right) _P}}\left( \rho \right) =\left\{ \left( \sigma
_{\left[ i_{\left( r\right) _P}\right] }\left( \rho \right) \right) ,\left(
\lambda _{\left[ i_{\left( r\right) _P}\right] }\left( \rho \right) \right)
\right\} \text{ (for all possible }\left[ i_{\left( r\right) _P}\right] )
\end{equation}
Here, we notice an interesting fact that every $CRF_{_{\left( r\right)
_P}}\left( \rho \right) ,$ as a matrix, is just equal to the partial trace $%
tr_{\left( r\right) _P}\left( \rho \right) \equiv tr_{r_1\cdots r_P}\left(
\rho \right) $ $.$ In fact, from the definition of the partial traces and
Eq.(5), this conclusion is obvious, however in the method of convex rigid
frames, $CRF_{_{\left( r\right) _P}}\left( \rho \right) $ always is regarded
as a CRF. Of course, for the distinct $\rho $ and $\rho ^{\prime }$,
generally, $CRF_{_{\left( r\right) _P}}\left( \rho \right) $ and $%
CRF_{_{\left( r\right) _P}}\left( \rho ^{\prime }\right) $ may be distinct.
In the following, for a fixed $\left( r\right) _P$, we use the symbol $%
CRF_{_{\left( r\right) _P}}\equiv \left\{ CRF_{_{\left( r\right) _P}}\left(
\rho \right) \mid :\rho \in {\Bbb P}_{M\times N}\ \right\} $ which is a set
of CRFs corresponding to various pure-states density matrixes $\rho $.

{\it Theorem 1.} For each fixed proper subset $\left( r\right) _P\equiv
\left\{ r_1,\cdots ,r_P\right\} \subset {\Bbb Z}_M,$ ($r_1<\cdots
<r_P,1\leqslant P\leqslant M-1),$ there is a 1-1 correspondence${\cal \ }%
T_{\left( r\right) _P}$ between the set ${\Bbb P}_{M\times N}$ and the set $%
CRF_{_{\left( r\right) _P}},$ symbolize this by $T_{\left( r\right) _P}:%
{\Bbb P}_{M\times N}\rightleftarrows CRF_{_{\left( r\right) _P}}$.

{\bf Proof. }As in the above, by using of{\bf \ }Eq.($5$) for every
pure-state $\rho $ there always is a corresponding $CRF_{_{\left( r\right)
_P}}\left( \rho \right) ,$ now we define the mapping $T_{\left( r\right) _P}$
by 
\begin{equation}
T_{\left( r\right) _P}:{\Bbb P}_{M\times N}\longrightarrow CRF_{_{\left(
r\right) _P}},T_{\left( r\right) _P}\left( \rho \right) =CRF_{_{\left(
r\right) _P}}\left( \rho \right) \text{ for }\rho \in {\Bbb P}_{M\times N}
\end{equation}
If $\rho ^{\prime }\neq \rho ,$ then $\mid \Psi >\neq \pm \mid \Psi >$, this
means that there is at least one of $N^P$ real numbers $\lambda _{\left[
i_{\left( r\right) _P}\right] }\left( \rho \right) $, or of $N^P$ matrixes $%
\sigma _{\left[ i_{\left( r\right) _P}\right] }\left( \rho \right) $ which
is different from one of $\lambda _{\left[ i_{\left( r\right) _P}\right]
}\left( \rho ^{\prime }\right) ,$ or of matrixes $\sigma _{\left[ i_{\left(
r\right) _P}\right] }\left( \rho ^{\prime }\right) ,$ thus $CRF_{_{\left(
r\right) _P}}\left( \rho \right) {}\neq CRF_{_{\left( r\right) _P}}\left(
\rho ^{\prime }\right) {}.$

Conversely, for any $CRF_{N^P}=\left\{ \left( \mu _k\right) ,\left( \omega
_k\right) \right\} \left( k=1,\cdots ,N^P\right) \in CRF_{_{\left( r\right)
_P}},$ we can take the set $\left\{ i_{r_1},\cdots ,i_{r_P}\right\} \left(
i_{r_1},\cdots ,i_{r_P}=0,\cdots ,N-1\right) $ to substitute the set $%
\left\{ i_1,\cdots ,i_P\right\} $ of indexes in the nature order of $\left\{
k\right\} ,$ and we can rewrite $CRF_{N^P}$ as $CRF_{N^P}=\left\{ \left( \mu
_{\left[ i_{\left( r\right) _P}\right] }\right) ,\left( \omega _{\left[
i_{\left( r\right) _P}\right] }\right) \right\} $. Suppose that the
pure-state $\mu _{\left[ i_{\left( r\right) _P}\right] }=\mid \xi \left[
i_{\left( r\right) _P}\right] ><\xi \left[ i_{\left( r\right) _P}\right]
\mid ,\mid \xi \left[ i_{\left( r\right) _P}\right]
>=\sum\limits_{i_{s_1},\cdots ,i_{s_{M-P}}=0}^{N-1}d_{i_{s_1}\cdots
i_{s_{M-P}}}\mid i_{s_1}\cdots i_{s_{M-P}}>,$ then we write $\mid \Phi
>=\sum\limits_{j_1,\cdots ,j_M=0}^{N-1}f_{j_1\cdots j_M}\mid j_1\cdots j_M>$
where $f_{j_1\cdots j_M}$ is determined by 
\begin{equation}
f_{i_1\cdots i_M}=\mu _{\left[ i_{\left( r\right) _P}\right]
}d_{i_{s_1}\cdots i_{s_{M-P}}}\text{ when as a set }\left( i_1\cdots
i_M\right) =\left( i_{r_1}\cdots i_{r_P}\right) \cup \left( i_{s_1}\cdots
i_{s_{M-P}}\right)
\end{equation}
It can be verified directly that $\mid \Phi >$ is a $M$-partite quNit
normalized pure-state, and we just have $T_{\left( r\right) _P}\left( \mid
\Phi ><\Phi \mid \right) =CRF_{N^P}$. $\square $

Since in the above discussion, $\left( r\right) _P$ and $\left( s\right)
_{M-P}$ are completely symmetric in status, thus by the similar way, for the
subset $\left( s\right) _{M-P}$ we have yet a $T_{\left( s\right) _{M-P}}:%
{\Bbb P}_{M\times N}\rightleftarrows CRF_{_{\left( s\right) _{M-P}}}$. From
the theorem 1, ${\Bbb P}_{M\times N}$ and the set $\left\{ CRF_{_{\left(
r\right) _P}}\right\} $ are 1-1 corresponding, therefore {\bf some studies
of the multipartite quNit pure-states can be returned into the studies about 
}$\left\{ CRF_{_{\left( r\right) _P}}\right\} $. In this Letter, we call
this way the {\bf method of convex rigid frames}. Sometimes, this method is
a more effective means. As the examples of applications, in the following we
use this method to study some basic problems.

The first is the partial separability problem. Generally, the common
so-called separability, in fact, is the `full-separability'. For the general
multipartite systems, the problem becomes even more complex. In fact, there
yet is other concept of separability weaker than full-separability, i.e. the
partial separability$,$ e.g. for a tripartite qubit pure-state $\rho _{ABC}$%
, there are the A-BC-separability, B-AC-separability, C-AB-separability,
etc.[2,3]. Related to Bell-type inequalities and some criteria of partial
separability of the multipartite systems, see [4-6]$.$

In the first place, we need to define strictly what is the partial
separability of a multipartite quNit pure-state. About this, we must
consider the order numbered by us of the particles. If two ordered proper
subsets $\left( r\right) _P\equiv \left\{ r_1,\cdots ,r_P\right\} \left(
\;1\leqslant r_1<\cdots <r_P\leqslant M\right) $\ and $\left( s\right)
_{M-P}\equiv \left\{ s_1,\cdots ,s_{M-P}\right\} \left( 1\leqslant
s_1<\cdots <s_{M-P}\leqslant M\right) $\ in ${\Bbb Z}_M$\ obey 
\begin{equation}
\left( r\right) _P\cup \left( s\right) _{M-P}={\Bbb Z}_M,\;\left( r\right)
_P\cap \left( s\right) _{M-P}=\emptyset
\end{equation}
where $P$\ is an integer, 1$\leqslant P\leqslant M-1,$\ then the set $%
\left\{ \left( r\right) _P,\left( s\right) _{M-P}\right\} $\ forms a
partition of ${\Bbb Z}_M$, in the following for the sake of stress, we
denote it by the symbol $\left( r\right) _P\Vert \left( s\right) _{M-P}$.
Now, for a given partition $\left( r\right) _P\Vert \left( s\right) _{M-P},$
we use the natural basis $\left\{ \mid i_{r_1}\cdots i_{r_P}i_{s_1}\cdots
i_{s_{M-P}}>\right\} $ and write 
\begin{equation}
\mid \Psi _{\left( r\right) _P\Vert \left( s\right) _{M-P}}>\equiv
\sum_{i_1,\cdots ,i_M=0}^{N-1}d_{i_1i_2\cdots i_M}\mid i_{r_1}\cdots
i_{r_P}i_{s_1}\cdots i_{s_{M-P}}>,\;d_{i_1i_2\cdots i_M}=c_{i_{r_1}\cdots
i_{r_P}i_{s_1}\cdots i_{s_{M-P}}}
\end{equation}
Obviously, $\mid \Psi _{\left( r\right) _P\Vert \left( s\right) _{M-P}}>$
and $\mid \Psi >$ in Eq.(1), in fact, are completely same in physic, the
difference only is the order numbered by us of particles. For instance, $%
\Psi _{A\Vert BCD}=\Psi _{AB\Vert CD}\;=\Psi _{ABC\Vert D}=\Psi _{ABCD}$,
and $\Psi _{AC\Vert BD}=\sum c_{ijkl}\mid i_Ak_Cj_Bl_C>,$ etc.. However, we
notice that, generally, $\rho _{\left( r\right) _P\Vert \left( s\right)
_{M-P}}\equiv \mid \Psi _{\left( r\right) _P\Vert \left( s\right)
_{M-P}}><\Psi _{\left( r\right) _P\Vert \left( s\right) _{M-P}}\mid \neq
\mid \Psi ><\Psi \mid =\rho $ under the standard basis $\left\{ \mid
i_1\cdots i_M>\right\} $, unless $\left( r\right) _P\Vert \left( s\right)
_{M-P}$\ just maintains the natural order of ${\Bbb Z}_M$\ (i.e. $\left(
r\right) _P=\left( 1,\cdots ,P\right) ,\left( s\right) _{M-P}=\left(
P+1,\cdots ,M\right) )$, so $\rho _{\left( r\right) _P\Vert \left( s\right)
_{M-P}}=\rho .$

{\it Definition 2.}{\bf \ }For the partition $\left( r\right) _P\Vert \left(
s\right) _{M-P}$ , a M-partite quNit pure-state $\mid \Psi >$ is called to
be $\left( r\right) _P-\left( s\right) _{M-P}$-separable$,$\ if the
corresponding $\mid \Psi _{\left( r\right) _P\Vert \left( s\right) _{M-P}}>$
\ can be decomposed as a product of two pure-states as 
\begin{equation}
\mid \Psi _{\left( r\right) _P\Vert \left( s\right) _{M-P}}>=\mid \Psi
_{\left( r\right) _P}>\otimes \mid \Psi _{\left( s\right) _{M-P}}>\text{ or }%
\rho _{\left( r\right) _P\Vert \left( s\right) _{M-P}}=\rho _{\left(
r\right) _P}\otimes \rho _{\left( s\right) _{M-P}}
\end{equation}
where $\mid \Psi _{\left( r\right) _P}>$ $\in \otimes _{\alpha
=1}^PH_{r_\alpha }$ ,$\rho _{\left( r\right) _P}=$\ $\mid \Psi _{\left(
r\right) _P}>$\ $<\Psi _{\left( r\right) _P}\mid $ and $\mid \Psi _{\left(
s\right) _{M-P}}>\in \otimes _{\alpha =1}^{M-P}H_{s_\alpha },\rho _{\left(
s\right) _{M-P}}=\mid \Psi _{\left( s\right) _{M-P}}><\Psi _{\left( s\right)
_{M-P}}\mid .$ If $\mid \Psi >$ \ is not $\left( r\right) _P-\left( s\right)
_{M-P}$-separable, then we call it to be $\left( r\right) _P-\left( s\right)
_{M-P}$-inseparable.

We notice that for the distinct partitions, $\rho $ can have distinct
partial separability. Of course, if a pure-state $\rho $ is partially
inseparable with respect to any partition, then it must be entangled.
Conversely, if a pure-state always is completely partially separable with
respect to all possible partitions $\left( r\right) _P\Vert \left( s\right)
_{M-P}$, then it is separable (disentangled, full-separable). By using the
above method of CRFs, we can obtain the following theorem, which, in fact,
is a geometric explanation of the partial separability of the $M$-partite
quNit pure-states.

{\it Theorem 2.} The sufficient and necessary conditions of the $M$-partite
quNit pure-state $\rho =\mid \Psi ><\Psi \mid $\ to be $\left( r\right)
_P-\left( s\right) _{M-P}$-separable is that CRF$_{_{\left( r\right) _P}}$\ $%
\left( \rho \right) $(or $CRF_{_{\left( s\right) _{M-P}}}\left( \rho \right) 
$) shrinks to one point (pure-state vertex), i.e. all d$\left( \sigma
_{\left[ i_{\left( r\right) _P}\right] }-\sigma _{\left[ i_{\left( r\right)
_P}^{\prime }\right] }\right) =0$ $($or all d$\left( \sigma _{\left[
i_{\left( s\right) _{M-P}}\right] }-\sigma _{\left[ i_{\left( s\right)
_{M-P}}^{\prime }\right] }\right) =0)$\ for any $i,i^{\prime }=0,\cdots ,N-1$%
.

{\bf Proof.} {\bf Necessity. }Suppose that the pure-state $\rho =\mid \Psi
><\Psi \mid $ is $\left( r\right) _P-\left( s\right) _{M-P}$-separable,
according to the Definition 2, this means that (see Eq. (9)) $\mid \Psi
_{\left( r\right) _P\Vert \left( s\right) _{M-P}}>=\mid \Psi _{\left(
r\right) _P}>\otimes \mid \Psi _{\left( s\right) _{M-P}}>.$ If the
normalized $\mid \Psi _{\left( r\right) _P}>$ and $\mid \Psi _{\left(
s\right) _{M-P}}>$ , respectively, are $\mid \Psi _{\left( r\right)
_P}>=\sum\limits_{i_{r_1},\cdots ,i_{r_P}=0}^{N-1}d_{i_{r_1}\cdots
i_{r_P}}\mid i_{r_1}\cdots i_{r_P}>$ and {\it \ }$\mid \Psi _{\left(
s\right) _{M-P}}>=\sum\limits_{i_{s_1},\cdots ,i_{s_{M-P}}=0}^{N-1}${\it e}$%
_{i_{s_1}\cdots i_{s_{M-P}}}\mid i_{s_1}\cdots i_{s_{M-P}}>$, then by a
direct calculation, in $CRF_{\left( r\right) _P}\left( \rho \right) $ we
have 
\begin{equation}
\lambda _{\left[ i_{\left( r\right) _P}\right] }\left( \rho \right) =\left|
d_{i_{r_1}\cdots i_{r_P}}\right| ^2\text{ and all }\sigma _{\left[ i_{\left(
r\right) _P}\right] }\left( \rho \right) =\mid \Psi _{\left( r\right)
_P}><\Psi _{\left( r\right) _P}\mid
\end{equation}
this means that $CRF_{\left( r\right) _P}\left( \rho \right) $ will indeed
shrink to a point. Similarly, for $CRF_{\left( s\right) _{M-P}}\left( \rho
\right) .$

{\bf Sufficiency.} If all $\sigma _{\left[ i_{\left( r\right) _P}\right]
}\left( \rho \right) $ shrink to a point $\sigma =\mid \varphi ><\varphi
\mid ,$ $\mid \varphi >=\sum\limits_{k_1,\cdots ,k_P=0}^{N-1}f_{k_1\cdots
k_P}\mid k_1\cdots k_P>,$ then according to Eqs.(9) and (10), this means
that $\mid \Psi _{\left( r\right) _P\Vert \left( s\right)
_{M-P}}>=\sum\limits_{i_{r_1},\cdots ,i_{r_P}}^{N-1}f_{k_1\cdots k_P}\mid
i_{r_1}\cdots i_{r_P}>\otimes \sum\limits_{i_{s_1},\cdots
,i_{s_{M-P}}}g_{i_{s_1}\cdots i_{s_{M-P}}}\mid i_{s_1}\cdots
i_{s_{M-P}}>=\mid \varphi >\otimes \mid \psi >,$ where $\mid \psi
>=\sum\limits_{i_{s_1},\cdots ,i_{s_{M-P}}}g_{i_{s_1}\cdots i_{s_{M-P}}}\mid
i_{s_1}\cdots i_{s_{M-P}}>,\ g_{i_{r_1}\cdots i_{r_P}}$ are some
coefficients, and $\mid k_1\cdots k_P>$ has been substituted by $\mid
i_{r_1}\cdots i_{r_P}>$. Therefore $\rho $ is $\left( r\right) _P-\left(
s\right) _{M-P}$-separable. $\square $

{\it Corollary.}{\bf \ }For a $M$-partite quNit pure-state $\rho ,$\ $%
CRF_{_{\left( r\right) _P}}\left( \rho \right) $\ and $CRF_{_{\left(
s\right) _{M-P}}}$\ $\left( \rho \right) $ \ both shrink to points, or both
not.

The proof is evident from the proof of the Theorem 2.

Therefore in view of method of CRFs, every separable multipartite quNit
(disentangled) pure-states is an extremely special state, i.e. of which all
CRFs must be shrunk to a point. As an simple example, we consider the
normalized tripartite qutrit pure-state $\rho _{3\times 3}=\mid \Psi
_{3\times 3}><\Psi _{3\times 3}\mid \in {\Bbb P}_{3\times 3}$,$\;\mid \Psi
_{3\times 3}>=\sum_{i,j,k=0}^2c_{ijk}\mid i_Aj_Bk_C>$ $\left(
\sum_{i,j,k=0}^2\left| c_{ijk}\right| ^2=1\right) $ and the partition $%
B\Vert AC,$ by a direct calculation we obtain a 3-CRF 
\begin{equation}
CRF_{\left( B\right) }\left( \rho _{3\times 3}\right) =\left\{ \left( \sigma
_{\left[ \left( j_B\right) \right] }\left( \rho _{3\times 3}\right) \right)
,\left( \lambda _{\left[ \left( j_B\right) \right] }\left( \rho _{3\times
3}\right) \right) \right\} (j=0,1,2)
\end{equation}
where $\lambda _{\left[ \left( j_B\right) \right] }\left( \rho _{3\times
3}\right) =\sum\limits_{i,k=0}^2\left| c_{ij_Bk}^2\right| ,\sigma _{\left[
\left( j_B\right) \right] }\left( \rho _{3\times 3}\right) =\mid \varphi
_{3\times 2}><\varphi _{3\times 2}\mid ,\mid \varphi _{3\times 2}>=\left(
\lambda _{\left[ \left( j_B\right) \right] }\right) ^{-\frac 12%
}\sum\limits_{i,k=0}^2c_{ij_Bk}\mid i_Ak_C>$. It is easily verified that the
condition $d\left( \sigma _{\left[ \left( j_B\right) \right] }\left( \rho
^{\left( 3\right) }\right) -\sigma _{\left[ \left( j_B^{\prime }\right)
\right] }\left( \rho ^{\left( 3\right) }\right) \right) =0\left( j,j^{\prime
}=0,1,2\right) $ leads to that for any $i,k=0,1,2$ all rates $%
c_{i0k}:c_{i1k}:c_{i2k}$ are equal, this is indeed the sufficient and
necessary conditions of $\mid \Psi ^{\left( 3\right) }>$ to be
B-AC-separable.

Secondly, we study the problem of classification of the $M$-partite quNit
pure-states. In view of the method of CRFs, a very natural way of
classification is to use the motions of the CRFs.

{\it Definition 3.} We call two $M$-partite quNit pure-states $\rho $\ and $%
\rho ^{\prime }$\ are `equivalent by motion', symbolize by $\rho \backsim
\rho ^{\prime }$, if and only if $CRF_{_{\left( r\right) _P}}\left( \rho
\right) $\ and $CRF_{_{\left( r\right) _P}}\left( \rho ^{\prime }\right) $\
are identical (see the Definition 1) with respect to all possible non-null
proper subset $\left( r\right) _P\left( 1\leqslant P\leqslant M-1\right) $.

A notable advantage of this definition is that this equivalence relation do
not break the partial separability of the $M$-partite quNit pure-states. In
fact, we have the following

{\it Corollary.\ }If two $M$-partite quNit pure-states $\rho $\ and $\rho
^{\prime }$\ are equivalent by motion, then $\rho $ \ and $\rho ^{\prime }$\
both are $\left( r\right) _P-\left( s\right) _{M-P}$-separable (or both $%
\left( r\right) _P-\left( s\right) _{M-P}$-inseparable), with respect to any 
$\left( r\right) _P\Vert \left( s\right) _{M-P},\;$i.e. the partial
separability is an invariant of class.

The proof is obvious.

Notice that about the above way of classification, we must still solve the
problem of reasonableness in physics, because in quantum information a
pure-states $\rho $, generally, represents some information status. It is
known that, generally, for the indistinguishability of multipartite quNit
states we must use the local operation and classical communications
(LOCC)[7-9]. Now we prove that our way is reasonable, i.e. we prove that the
above classification by motions, in fact, is just the classification of $%
{\Bbb P}_{M\times N}$ under the local unitary transformations (LUs). In
order to prove this, in fact, we only need to prove the following theorem.

{\it Theorem 3.} Two $M$-partite quNit pure-states $\rho $\ and $\rho
^{\prime }$\ are equivalent by motion (see the Definition 2), if and only if
there are M unitary matrixes $u_i\left( N\right) \in U\left( N\right) \left(
i=1,\cdots ,M\right) $ that ${\it the}$%
\begin{equation}
\rho ^{\prime }=u_1\left( N\right) \otimes \cdots \otimes u_M\left( N\right)
\rho u_M^{\dagger }\left( N\right) \otimes \cdots \otimes u_1^{\dagger
}\left( N\right)
\end{equation}
{\it \ }

{\bf Proof.} In the first place, we notice that in the H-S space only the
unitary transformations of operators can keep the invariances of distances
and modulus of the vectors, and a tensor product of some unitary matrixes is
still a unitary matrix. Now, if Eq.($13$) holds, according to Eqs.(2,3,5),
the change from $CRF_{_{\left( r\right) _P}}\left( \rho \right) $ to $%
CRF_{_{\left( r\right) _P}}\left( \rho ^{\prime }\right) $ for each $\left(
r\right) _P$ is determined by a unitary matrix $u_{s_1}\left( N^{M-P}\right)
\otimes \cdots \otimes u_{s_{M-P}}\left( N^{M-P}\right) $, which acts upon
every `part' $\mid \Psi \left[ i_{\left( r\right) _P}\right] >$ of $\mid
\Psi >$ in Eq.(2) and keeps $\eta _{\left[ i_{\left( r\right) _P}\right]
}\left( \rho \right) $ to be invariant$,$ thus the identical relation
between $CRF_{_{\left( r\right) _P}}\left( \rho \right) $ and $CRF_{_{\left(
r\right) _P}}\left( \rho ^{\prime }\right) $ is quite obvious.

Conversely, since $\lambda _{\left[ i_{\left( r\right) _P}\right] }\left(
\rho \right) =\lambda _{\left[ i_{\left( r\right) _P}\right] }\left( \rho
^{\prime }\right) $ for all possible $\left[ i_{\left( r\right) _P}\right] $%
, we know that for every $\left[ i_{\left( r\right) _P}\right] $ there must
be a unitary matrix $u_{\left[ i_{\left( r\right) _P}\right] }\left(
N^{M-P}\right) $ which acts upon the `part' $\mid \Psi \left[ i_{\left(
r\right) _P}\right] >$ of $\mid \Psi >$, and keeps that all equations $%
d\left( \sigma _{\left[ i_{\left( r\right) _P}\right] }\left( \rho \right)
,\sigma _{\left[ i_{\left( r\right) _P}^{\prime }\right] }\left( \rho
\right) \right) =d\left( \sigma _{\left[ i_{\left( r\right) _P}\right]
}\left( \rho ^{\prime }\right) ,\sigma _{\left[ i_{\left( r\right)
_P}^{\prime }\right] }\left( \rho ^{\prime }\right) \right) $ always hold$.$
This fact must hold for arbitrary $\left( r\right) _P$ and arbitrary set $%
\left[ i_{\left( r\right) _P}\right] $ of indexes$,$ the unique possibility
is that there are some $u_1\left( N\right) ,\cdots ,u_M\left( N\right)
,u_k\left( N\right) \in U\left( N\right) \left( k=1,\cdots ,M\right) $ and $%
\mid \Psi ^{\prime }>=u_1\left( N\right) \otimes \cdots \otimes u_M\left(
N\right) \mid \Psi >.$ $\square $

This theorem gives us a perfect explanation of the LUs, i.e. a LU acting
upon $\rho $, in fact, is a motion of CRFs, as a motion of a rigid body as
in the classical mechanics.

Thirdly, we discuss the invariants of the classification and a possible
explanation. Since the distance between two points (vectors) in the H-S
space is invariant under any motion (LU), evidently there are at least two
kinds of invariants of motions (LUs): The the volumes of the convex
polyhedrons propped up by the CRFs, others are the angles of intersections
of any two `props' in every CRF.

As for the problem how to calculate the volume of a convex polyhedron in the
H-S space, see [10] and its references. For a $M$-partite quNit pure-state $%
\rho $ and a given $\left( r\right) _P\Vert \left( s\right) _{M-P},$ let $%
V\left( CRF_{_{\left( r\right) _P}}\left( \rho \right) \right) $ denote the
volumes of the convex polyhedron with $N^P$ vertexes $\left\{ \sigma
_{\left[ i_{\left( r\right) _P}\right] }\right\} $(for all possible $\left[
i_{\left( r\right) _P}\right] )$ as in Eq.(5). Similarly, $V\left(
CRF_{_{\left( s\right) _{M-P}}}\left( \rho \right) \right) .$ Now we denote
the pair of volumes by 
\begin{equation}
V_{\left( r\right) _P\Vert \left( s\right) _{M-P}}\left( \rho \right)
=\left[ V\left( CRF_{_{\left( r\right) _P}}\left( \rho \right) \right)
,V\left( CRF_{_{\left( s\right) _{M-P}}}\left( \rho \right) \right) \right]
\end{equation}
Obviously, $V_{\left( r\right) _P\Vert \left( s\right) _{M-P}}\left( \rho
\right) $ is a invariant under motions (LUs) of $\rho $. In addition, in $%
CRF_{_{\left( r\right) _P}}\left( \rho \right) ,$ the direction from the
point $\sigma _{_{\left( r\right) _P}}\left( \rho \right) =\left\{ \lambda
_{\left[ i_{\left( r\right) _P}\right] }\left( \rho \right) \right\} $ (for
all possible $\left[ i_{\left( r\right) _P}\right] )$ to a fixed vertex $%
\sigma \left[ k_{\left( r\right) _P}\right] \left( k=0,\cdots ,N-1\right) $
of $CRF_{_{\left( r\right) _P}}\left( \rho \right) $ can be expressed by the
vector 
\begin{equation}
\omega _{_{\left[ k_{\left( r\right) _P}\right] }}\left( \rho \right) =\sum_{%
\text{for all possible }\left[ i_{\left( r\right) _P}\right] }\left( \lambda
_{\left[ i_{\left( r\right) _P}\right] }\left( \rho \right) -\delta _{\left[
i_{\left( r\right) _P}\right] ,\left[ k_{\left( r\right) _P}\right] }\right)
\sigma _{\left[ i_{\left( r\right) _P}\right] }\left( \rho \right)
\end{equation}
Therefore the angle (we label it by $\theta \left( \rho ,\left[ k_{\left(
r\right) _P}\right] ,\left[ l_{\left( r\right) _P}\right] \right) )$ of
intersection of two directs $\omega _{\left[ k_{\left( r\right) _P}\right]
}\left( \rho \right) $ and $\omega _{\left[ l_{\left( r\right) _P}\right]
}\left( \rho \right) $ can be determined by 
\begin{equation}
\cos \theta \left( \left[ k_{\left( r\right) _P}\right] ,\left[ l_{\left(
r\right) _P}\right] ,\rho \right) =<\omega _{\left[ k_{\left( r\right)
_P}\right] }\left( \rho \right) ,\omega _{\left[ l_{\left( r\right)
_P}\right] }\left( \rho \right) >\diagup \left\| \omega _{\left[ k_{\left(
r\right) _P}\right] }\left( \rho \right) \right\| \cdot \left\| \omega
_{\left[ l_{\left( r\right) _P}\right] }\left( \rho \right) \right\|
\end{equation}
Obviously, $\cos \theta \left( \left[ k_{\left( r\right) _P}\right] ,\left[
l_{\left( r\right) _P}\right] ,\rho \right) $ is yet an invariant under
motions (LUs) of $\rho $.

At present, we cannot yet understand what is the meaning of $\cos \theta
\left( \rho ,\left[ k_{\left( r\right) _P}\right] ,\left[ l_{\left( r\right)
_P}\right] \right) $ in quantum information. However, we find a quite
natural explanation of $V_{\left( r\right) _P\Vert \left( s\right)
_{M-P}}\left( \rho \right) $ as follows. From the Theorem 2, its corollary
and the fact that a convex polyhedron shrinks to a point if and only if its
volume vanishes, then we know that $\rho $ is $\left( r\right) _P-\left(
s\right) _{M-P}$-separable if and only if $V_{\left( r\right) _P\Vert \left(
s\right) _{M-P}}\left( \rho \right) =\left( 0,0\right) .$ Conversely, if $%
V_{\left( r\right) _P\Vert \left( s\right) _{M-P}}\left( \rho \right) \neq
\left( 0,0\right) $ then $\rho $ is $\left( r\right) _P-\left( s\right)
_{M-P}$-inseparable, where the value of $V\left( CRF_{_{\left( r\right)
_P}}\left( \rho \right) \right) $ means that the degree of the difficulty of
the factor $\rho _{\left( r\right) _P}$ to be separated out from $\rho .$
Similarly, for $V\left( CRF_{_{\left( s\right) _{M-P}}}\left( \rho \right)
\right) .$ Therefore we can regard that $V_{\left( r\right) _P\Vert \left(
s\right) _{M-P}}\left( \rho \right) $ denotes the degree of the measure of
the $s_{\left( r\right) _P}-s_{\left( s\right) _{M-P}}$-inseparability. It
is quite interesting that, generally, $V\left( CRF_{_{\left( r\right)
_P}}\left( \rho \right) \right) \neq V\left( CRF_{_{\left( s\right)
_{M-P}}}\left( \rho \right) \right) $ unless they both vanish, this means
that the above two degrees of the difficulties can be different. In
addition, it is known that if $\rho $ is $\left( r\right) _P-$ $\left(
s\right) _{M-P}$-inseparability, then there must be the so-called partial
entanglement[4,6]. What a pity, an entanglement measure, generally, should
be in form as a Neumann entropy and it must at least obey some limits[$11$],
however $V_{\left( r\right) _P\Vert \left( s\right) _{M-P}}\left( \rho
\right) $ has no these properties, so we cannot taken it as a measure of the
partial entanglement.


\begin{references}
\bibitem{}  V. Vedral and M. B.. Plenio, Phys. Rev. A, {\bf 57}(1998)1619.

\bibitem{}  W. D\"{u}r, G. Vidal, and J. I. Cirac, Phys. Rev. A, {\bf 62}%
(2000)062314.

\bibitem{}  F. Verstraete, J. Dehaene, B. De Moor, and H. Verschelde, Phys.
Rev. A, {\bf 65}(2002)052112.

\bibitem{}  M. Seevinck and G. Svetlichny, Phys. Rev. Lett., {\bf 89}%
(2002)060401.

\bibitem{}  A. O. Pittenger and M. H. Rubin, Phys. Rev. A, {\bf 62}%
(2000)032313.

\bibitem{}  T. Yamakami, quant-ph/0308072.

\bibitem{}  C. H. Bennett, D. P. DiVincenzo, J. Smolin, and W. K. Wootters,
Phys. Rev. A, {\bf 54}(1996)3824.

\bibitem{}  C. H. Bennett, H. J. Bernstein, S. Popescu, and B. Schumacher,
Phys. Rev. A, {\bf 53}(1996)2046.

\bibitem{}  P. Hayden, B. M. Terhal and A. Uhlmann, Phys. Rev. Lett., {\bf 86%
}(2001)5807.

\bibitem{}  K. Zyczkowski, and H. J. Sommers, J. Phys. A, {\bf 36}%
(2003)10115.

\bibitem{}  M. Horodecki, P. Horodecki and R. Horodecki, Phys. Rev. Lett., 
{\bf 84}(2000)2014.
\end{references}
\end{document}